\documentclass[runningheads]{llncs}

\usepackage{ enumitem, dirtytalk, pgfplots, caption, xcolor, graphicx, booktabs, amssymb, appendix, listings, bchart, hyperref, subcaption}
\usetikzlibrary{patterns}

\definecolor{codegreen}{rgb}{0,0.6,0}
\definecolor{codegray}{rgb}{0.5,0.5,0.5}
\definecolor{codepurple}{rgb}{0.58,0,0.82}
\definecolor{backcolour}{rgb}{0.95,0.95,0.92}

\lstdefinestyle{lst_style}{
    backgroundcolor=\color{backcolour},   
    commentstyle=\color{codegreen},
    keywordstyle=\color{magenta},
    numberstyle=\tiny\color{codegray},
    stringstyle=\color{codepurple},
    basicstyle=\ttfamily\scriptsize,
    breakatwhitespace=false,         
    breaklines=true,                 
    captionpos=b,                    
    keepspaces=true,                 
    numbers=none,                    
    numbersep=5pt,                  
    showspaces=false,                
    showstringspaces=false,
    showtabs=false,                  
    tabsize=2,
    frame=single
}

\colorlet{punct}{red!60!black}
\definecolor{background}{HTML}{EEEEEE}
\definecolor{delim}{RGB}{20,105,176}
\colorlet{numb}{magenta!60!black}

\lstdefinelanguage{json}{
    basicstyle=\normalfont\ttfamily,
    numbers=left,
    numberstyle=\scriptsize,
    stepnumber=1,
    numbersep=8pt,
    showstringspaces=false,
    breaklines=true,
    backgroundcolor=\color{background},
    literate=
     *{0}{{{\color{numb}0}}}{1}
      {1}{{{\color{numb}1}}}{1}
      {2}{{{\color{numb}2}}}{1}
      {3}{{{\color{numb}3}}}{1}
      {4}{{{\color{numb}4}}}{1}
      {5}{{{\color{numb}5}}}{1}
      {6}{{{\color{numb}6}}}{1}
      {7}{{{\color{numb}7}}}{1}
      {8}{{{\color{numb}8}}}{1}
      {9}{{{\color{numb}9}}}{1}
      {:}{{{\color{punct}{:}}}}{1}
      {,}{{{\color{punct}{,}}}}{1}
      {\{}{{{\color{delim}{\{}}}}{1}
      {\}}{{{\color{delim}{\}}}}}{1}
      {[}{{{\color{delim}{[}}}}{1}
      {]}{{{\color{delim}{]}}}}{1},
}

\pgfplotsset{compat=1.17}

\begin{document}

\title{
    Accept All: The Landscape of Cookie Banners in Greece and the UK%
    \thanks{This is an accepted manuscript of a paper by the same title to appear in the proceedings of IFIP SEC 2021: Conf.\ on ICT Systems Security and Privacy Protection.}
}

\author{
    Georgios Kampanos \and Siamak F. Shahandashti%
}

\authorrunning{G. Kampanos and S. F. Shahandashti}

\institute{
    University of York, UK\\ 
    \email{kampanosg@outlook.com}, 
    \email{siamak.shahandashti@york.ac.uk} 
}

\maketitle

\begin{abstract}
Cookie banners are devices implemented by websites to allow users to manage their privacy settings with respect to the use of cookies. They are part of a user's daily web browsing experience since legislation in Europe requires websites to show such notices. In this paper, we carry out a large-scale study of more than 17,000 websites including more than 7,500 cookie banners in Greece and the UK to determine compliance and tracking transparency levels. Our analysis shows that although more than 60\% of websites store third-party cookies in both countries, only less than 50\% show a cookie notice and hence a substantial proportion do not comply with the law even at the very basic level. We find only a small proportion of the surveyed websites providing a direct opt-out option, with an overwhelming majority either nudging users towards privacy-intrusive choices or making cookie rejection much harder than consent. Our results differ significantly in some cases from previous smaller-scale studies and hence underline the importance of large-scale studies for a better understanding of the big picture in cookie practices. 
\end{abstract}

\section{Introduction}
Websites implement cookie banners to allow users to either consent to or reject third-party cookie tracking and manage their privacy settings. After tighter legislation came into force, namely EU's General Data Protection Regulation (GDPR) and UK's Data Protection Act 2018 (DPA), more and more websites have adopted such notices, making cookie banners a part of users' everyday life. 

In theory, cookie banners (a.k.a. cookie notices) exist to empower users by informing them about tracking activity and allowing them to opt out if they wish to. However, real-world implementations of cookie banners appear to be a nuisance more than anything else~\cite{kulyk2018website}. Many websites design their notices to make opting out extremely hard, or remove the option completely as previous studies found~\cite{nouwens2020dark}. Furthermore, pre-selected options that nudge users towards privacy-intrusive choices are rife and significantly impact user behaviour~\cite{utz2019informed}. 
Both EU and UK regulators have clearly identified such  practices non-compliant with the GDPR and the DPA, including consent not being explicit and cookie rejection not being as easy as acceptance (See e.g.\ the EU's 2002 ePrivacy Directive~\cite{eprivacy-directive-2002}, the 2020 European Data Protection Board guidelines~\cite{edpb-guidelines-2020}, and discussions by Nouwens et al.~\cite{nouwens2020dark}). 
Yet, even if users manage to navigate around the maze of options and select their privacy settings, their choices are more likely to be ignored entirely as a study of Consent Management Providers (CMPs) deployed in European websites observed~\cite{matte2020cookie}. Worryingly, we have seen such \say{dark patterns} employed by big-tech, such as Facebook~\cite{council2018deceived}. 

The insight we have about the cookie banner landscapes and how they have changed as a result of legislation is mainly based on the analyses carried out on samples of high-traffic websites. Although such studies provide valuable information on how popular websites implement cookie banners, a natural question to ask is how well such observations generalise if a more comprehensive sample including lower-traffic websites is analysed. 
In this work, we aim to take a step towards investigating this question in the UK and Greece web landscapes. 

We set out to establish the types of cookie banners with which users have to interact  on a daily basis. 
Moreover, we will explore the distribution and availability of choices provided to users through cookie banner implementations. 
Using purpose-built software and with the aid of OpenWPM~\cite{englehardt2016census}, we collected, categorised and analysed more than 7,500 cookie banners from more than 17,000 websites across Greece and the UK. We discuss our findings which interestingly in some cases substantially differ with previous results in the literature. Our results therefore is a step towards developing a more clear and comprehensive understanding of the cookie banner landscape in the two countries. 

We consider Greece and the UK because of our familiarity with the respective languages and our hope that the comparison between the two provides interesting insight. On the one hand, websites in both countries adhere to very similar data protection laws. On the other hand however, the two countries vastly differ in their population size and their citizens' use of internet services~\cite{desi-report-2020}. 

\section{Related Work}
Studies in this area have mainly focused on the prevalence of cookie banners, the type of privacy options they offer, and whether they comply with the law.

In 2018, The Norwegian Consumer Council reviewed whether user interfaces of cookie notices and privacy settings provided by Google, Facebook and Microsoft Windows 10 discourage users from making privacy-aware choices~\cite{council2018deceived}. They found that all three companies offer default settings that are considered privacy intrusive, and that the cookie notices contain misleading wording while privacy-friendly options require multiple steps to find. 
They noted that Google and Facebook \say{threaten users with loss of functionality or deletion of the user account} unless they agree to those privacy-intrusive settings.

A number of studies in this area focus on providing a big picture across the world or Europe. 
Habib et al. conducted a 150-website analysis in 2018--19 and found that privacy options are frequent within their sample with 89\% websites with targeted advertising offering a way to opt-out~\cite{habib2019empirical}. However, they observed that, when visited from the US, only 28 out of 150 websites they considered displayed a cookie banner with only 5 of them offering a means to opt out. 
Degeling et al.'s study of 6,759 websites across the EU found adoption of cookie banners across the EU go up from 46.1\% before the GDPR to 62.1\% afterwards~\cite{degeling2018we}. 
Utz et al. carried a manual inspection of 1,000 popular websites in the EU and observed that 27.8\% provide no options, 68.0\% allow confirmation only, while only 3.2\% give a binary accept/reject choice~\cite{utz2019informed}. 
Another study of top 100 websites in each EU country by van Eijk et al. found 52\% of UK and 29\% of Greek websites implementing a cookie banner~\cite{eijk2019impact}. 

Two recent studies have looked at whether cookie banners provided by Content Management Platforms (CMPs) adhere to EU regulations. 
In a study published in 2019, Matte, Bielova and Santos surveyed 1,427 European websites from which they found that 141 websites registered an affirmative consent before the user had performed any actions and 38 websites offered no \say{opt-out} option at all~\cite{matte2020cookie}. The authors observed that at least 50\% of the websites in their dataset had pre-selected privacy options and at least 27 websites did not respect the user’s choice even though they declined to be tracked by cookies. 
In a study published in 2020 considering 680 websites, Nouwens et al. found that 32\% of them assumed \say{implicit consent} (agreeing without having any other option)~\cite{nouwens2020dark}, which make those websites not-compliant with GDPR. They also found that only 13\% of websites had a \say{reject all} button which almost always required additional clicks to be seen by a user.

Taking a closer look at the sample sizes of the studies that focus on providing a big picture, we have Habib et al.'s study of 150 websites worldwide (50 from each group of high, middle, and low popularity)~\cite{habib2019empirical}, Degeling et al.'s sample of 6,759 websites including 463 UK and 443 Greek ones~\cite{degeling2018we}, Utz et al.'s 1,000 randomly chosen from 500 top-ranking in each EU country~\cite{utz2019informed}, and van Eijk et al.'s sample of top 100 in each EU country~\cite{eijk2019impact}. 
Similarly for the studies that limit their attention to websites with CMP-provided banners, these include Nouwens et al.'s study of 680 such websites in the UK~\cite{nouwens2020dark} and Matte, Bielova, and Santos's investigation of 1,426 websites including 149 \texttt{.uk} and 53 \texttt{.gr} websites~\cite{matte2020cookie}. 
Although such studies provide valuable insight, none goes beyond 700 websites in the two countries we consider. This is understandable due to the complexity of automating such studies and that the goal of the said studies was to focus on a global view or on CMPs, and not on the comprehensive landscape in specific countries. 
This opens a natural question whether similar trends can be seen if the scale of the sample sizes considered are increased. Indeed, it is not clear whether characteristics observed in high-traffic websites remain similar if low-traffic ones are considered. 
To answer this question, we focused on two specific countries: the UK and Greece, but scaled up the sample size nearly ten-fold by automating our scraping and analysis, allowing us to expand our research to more than 14,000 UK and 3,000 Greek websites. 

While we developed purpose-built software, this study relied heavily upon existing software. We extended OpenWPM, an open-source web privacy measurement tool developed by Englehardt and Narayanan in 2016 to scrape and collect data~\cite{englehardt2016census}. OpenWPM allows researchers to detect and measure the use of third-party cookies (TPs), cookie synchronisation, as well as fingerprinting techniques. We modified OpenWPM to be able to recognise and store the website's cookie banner if one exists. 

\section{Research Questions, Methodology, and Implementation}

In light of the results of the previous works, we aim to investigate the following research questions through a large-scale study of Greek and UK domains: 

\begin{enumerate}[label=\textbf{RQ\arabic*}:, ref=RQ\arabic*, leftmargin=1.05cm]
    \item \label{rq:prevalence} What is the prevalence of cookie banners across the board when less popular websites in Greece and the UK are also considered? 
    \item \label{rq:avg_options} How does the distribution of options offered in cookie banners look like and what proportion of websites provide a direct cookie rejection option?
    \item \label{rq:no_options} What proportion of websites employ implicit consent?
    \item \label{rq:manage_options_count} What proportion of cookie banners allows their users to manage their privacy settings and control which vendors track them?
    \item \label{rq:distribution} How do the countries compare in terms of the privacy options offered by cookie banners? 
\end{enumerate}

Our data collection included three steps. We first built a comprehensive set of functioning websites to analyse and extract their cookie banners. Then we crawled the identified websites and collected relevant data such as the source code of the cookie notices and screenshots of the webpages. Finally, we sanitised and structured the collected data into a data structure that facilitates analysis. 
In the following, we explain these steps in more detail. 
The code developed for this study and referred to throughout the paper is publicly available at the following repository: \url{https://github.com/kampanosg/i-like-cookies}. 

\subsection{Building the Target List}
The first step is to identify websites to be analysed in this study. Using the Tranco top sites ranking~\cite{LePochat2019}, popular websites for the two countries were identified based on their Top Level Domains (TLDs): \texttt{.uk} and \texttt{.gr}. 
We decided to augment the TLD-based lists with other curated country-specific lists since many websites do not use the TLD of their country of origin, e.g.,  British Airways uses \texttt{.com}. 

\paragraph{Ethical Considerations.}
Not all websites allow crawling and many explicitly state that they only allow \say{personal use} of their services and content to their visitors. To respect such restrictions imposed by the websites, we developed two parsers to identify and exclude such websites from our automated crawl: 

\begin{enumerate}
    \item A Robots Exclusion Standard parser (\texttt{step1b\_checker\_robots.py}) that verifies whether websites allow crawling by reading their \texttt{robots.txt} file;
    
    \item A Terms of Service (ToS) parser (\texttt{step1c\_checker\_tos.py}) that makes best effort to find exclusionary terms, e.g. \say{for personal use only}, to comply with the ToS of each website.
\end{enumerate}

\subsection{Collecting Cookie Banners}
The second step is to effectively identify and collect the cookie banners on the compiled set of websites. We achieve this by taking advantage of the \say{I don’t care about cookies} (IDCAC) list (\url{www.i-dont-care-about-cookies.eu}), which provides an extensive selection of standard CSS selectors that cookie banners use. We parse these selectors and add them to a database. Furthermore, during testing, we identified and added 64 additional selectors to IDCAC.

After setting up the cookie selectors database, OpenWPM uses it to identify the cookie banners within the visited websites. We extended OpenWPM to detect cookie banners within a given website. For each website, we check whether it contains a CSS selector from the cached IDCAC list, using Selenium (\url{www.selenium.dev}), which allows for searching the HTML code of the website. When OpenWPM identifies a cookie banner, we perform additional analysis to make sure it is not a false positive, briefly by first ensuring that Selenium returned a valid HTML of a reasonable length, and then verifying that the returned HTML contains the terms \say{cookie} or \say{cookies}. 

When OpenWPM successfully identifies a cookie banner, it is stored in a database for further analysis. Combining the cached cookie selectors and Selenium's efficient Document Object Model (DOM) search enables the cookie banner extension to be efficient and robust. 

\subsection{Classifying and Normalising the Data}
To our knowledge, no standard exists for cookie banners, and therefore, every website has a different implementation for their notices. Thus, the HTML code and the options they provide can be drastically different from website to website. Such complexity can make the data analysis difficult. Thus, before performing any research on the data, we transformed them into a consistent data structure. 

\begin{table}[t]
    \centering
    \caption{The developed cookie banner options categories.}
    \begin{tabular}{@{}l@{\quad}l@{}}
    \toprule
        \textbf{Category} & \textbf{Description}                                   \\ \midrule
        Affirmative     & Options that prompt users to accept the use of cookies, \\
                        & e.g. \say{accept}, \say{agree}, \say{allow}, and \say{OK}.                                  \\
        Negative        & Options that allow users to opt-out from cookie tracking,        \\ 
                        & e.g. \say{decline}, \say{reject}, \say{disagree}, and \say{no}.                            \\
        Informational   & Options that take users to informational pages, e.g. \\
                        & \say{Privacy Policy}, \say{learn/see more}, and \say{see/show details}.                               \\
        Managerial      & Options that allow users to opt in/out of specific trackers,  \\ 
                        & e.g. \say{manage}, \say{settings}, and \say{vendors/partners}.                     \\ \bottomrule
    \end{tabular}
    \label{tab:privacy_options_categories}
\end{table}

\begin{figure}[t]
    \centering
    \includegraphics[width=\textwidth]{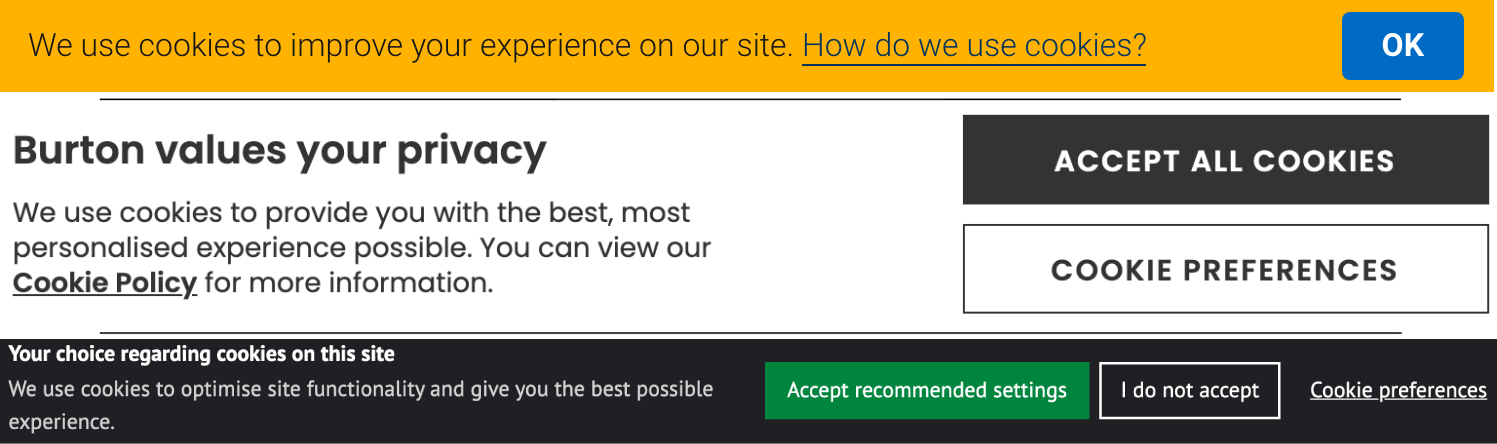}
    \caption{Three examples of cookie banners with different privacy options. 
             Top: Affirmative and Informational, 
             Middle: Affirmative, Managerial, and Informational, 
             Bottom: Affirmative, Negative, and Managerial.}
    \label{fig:cookie_banners}
\end{figure}

First, we sanitised the collected data to identify and then classify the privacy options within the collected cookie banners. We identified four privacy option categories: \emph{Affirmative}, \emph{Negative}, \emph{Informational} and \emph{Managerial} as defined in Table \ref{tab:privacy_options_categories}. We developed these four categories by manually inspecting a random sample of the collected data during testing, further informed by our own experience with cookie banners in the wild. Using these categories allows us to classify the cookie banners by the types of options they provide and hence better understand user choices. Examples of cookie banners providing different combinations of these options can be seen in Fig.~\ref{fig:cookie_banners}.

Manual inspection of the privacy options was necessary to account for local nuances. For example, although the noun $\alpha\pi o\delta o\chi\eta$ (lit. acceptance) was the most popular Affirmative call to action, a large number of banners used the verb $\delta\epsilon\chi o\mu\alpha\iota$ (I accept) indicating the same. We developed a comprehensive list of such variations and wrote a Python script (\texttt{step3a\_parse\_cookie\_banners.py}) to categorise cookie banners.

After we categorised the privacy options, we transformed the collected data into a consistent data structure that allows for efficient querying. More specifically, we converted the arbitrary HTML form of the collected cookie notices into JSON to facilitate both manual and automated analyses. 
Fig.~\ref{fig:html2json} depicts an example banner before and after categorisation and normalisation.

\begin{figure}[t]
    \centering
    \begin{lstlisting}[language=HTML, style=lst_style]
    <div class="..." aria-hidden="true" data-contents="...">
     <div class="container">
      <p class="...">By visiting x.gr you agree to the use of cookies.</p>
      <a href="#" class="...">Agree</a>
     </div>
    </div>
    \end{lstlisting}
    \begin{lstlisting}[language=json, style=lst_style]
    {
     "has_accept_btn": 1, 
     "accept-btn_cta": "Accept", 
     "privacy_text": "By visiting x.gr you agree to the use of cookies.", 
     ... 
    }
    \end{lstlisting}
    \caption{A cookie banner before (top) and after (bottom) normalisation}
    \label{fig:html2json}
\end{figure}

\section{Data and Results}

In this section we specify the dataset and discuss our findings. 

\subsection{The Collected Dataset}
\label{sec:dataset}

\paragraph{Websites.}
The Tranco list contains a total of 1\,M websites. From there, 3,446 are \texttt{.gr} websites and 18,768 are \texttt{.uk} ones. The additional country-specific lists provided an additional 674 websites: 40 additional Greek websites from TopGR (\url{https://topgr.gr}) and 634 additional UK websites from Kadaza (\url{www.kadaza.co.uk}) and Finder (\url{www.finder.com/uk}). Furthermore, we removed 125 Greek and 305 UK websites from the dataset as they were not accessible. In total, the initial dataset contained 3,361 Greek and 19,097 UK available websites. 

We checked each website to determine whether they allow crawlers. The Robots Exclusion Standard parser yielded 3,157 Greek (93\%), and 15,410 UK (69\%) websites that allowed crawling. Then the Terms of Service parser determined that 3,087 Greek (91\%) and 14,650 UK (65\%) websites permitted our study to crawl them. Table~\ref{tab:data_websites}, summarises the breakdown of our dataset.

\begin{table}[t]
    \centering
    \caption{Breakdown of the number of websites per country that are included (by source) and excluded (by reason)}
    \begin{tabular}{@{}l@{\quad}r@{\quad}r@{\quad}r@{}}
    \toprule
        & \textbf{Greece}   & \textbf{UK}       & \textbf{Combined} \\ 
    \midrule
    Included from Tranco 
        & 3,446             & 18,768            & 22,214            \\
    Included from country-specific lists 
        & 40                & 634               & 674               \\
    Excluded due to unavailability 
        & $-$125            & $-$305            & $-$430              \\

    Excluded due to \texttt{robots.txt} 
        & $-$204            & $-$3,687          & $-$3,891          \\
    Excluded due to Terms of service 
        & $-$70             & $-$760            & $-$830            \\
    \midrule 
    \textbf{Total studied} 
        & \textbf{3,087}    & \textbf{14,650}   & \textbf{17,737}   \\ 
    \bottomrule
    \end{tabular}
    \label{tab:data_websites}
\end{table}

\paragraph{OpenWPM.}

Successfully crawling websites for their cookie banners was possible by extending OpenWPM. In addition to cookie banners, OpenWPM also collected  more than 15\,M data points in Greece and the UK. This included information about the HTTP Requests and Responses (3.9\,M), scripts that a website loads (7.6\,M), and cookies stored in a user's web browser (2.3\,M). 

\paragraph{Viking.}
Collecting cookie banners for thousands of websites is a highly computing-intensive task, requiring over 24~hours for a complete crawl in Greece for example, even with parallel crawlers. To overcome this limitation we utilised University of York's Viking cluster\footnote{See \url{www.york.ac.uk/it-services/research-computing/viking-cluster}}, a high-performance computing cluster with 173 nodes, 42\,TB of memory, and 7024 Intel cores. While only using a fraction of Viking's resources (128\,GB of memory and 32 cores) the crawl was completed in 8~hours for Greece and just over 36~hours for the UK.

\subsection{Findings}
\subsubsection{\ref{rq:prevalence}: Prevalence depends on sample size.}
Our findings show that almost half of the websites we surveyed display a cookie banner. More specifically, around 48\% of Greek and 44\% of the UK websites included a cookie notice. 

\begin{table}[t]
    \centering
    \caption{Comparison of measured cookie banner prevalence rates. Sample sizes are approximates. Ranges indicate two methods of measurement.}
    \begin{tabular}{@{}l@{\quad}c@{\quad}c@{\quad}c@{\quad}c@{\quad}c@{}}
        \toprule
        \textbf{Study} & \textbf{Year} & \multicolumn{2}{c}{\textbf{Sample Size}} & \multicolumn{2}{c}{\textbf{Prevalence}} \\ 
        & \textbf{Conducted} & \textbf{UK} & \textbf{GR} & \textbf{UK} & \textbf{GR} \\ 
        \midrule 
        Degeling et al.~\cite{degeling2018we} & 2018 & 500 & 500 & 67--82\% & 60--69\% \\ 
        van Eijk et al.~\cite{eijk2019impact} & 2019 & 100 & 100 & 52\% & 29\% \\ 
        This work & 2020 & 14,000 & 3,000 & 44\% & 48\% \\ 
        \bottomrule
    \end{tabular}
    \label{tab:prev-comp}
\end{table}

When comparing our results with previous works of Degeling et al.~\cite{degeling2018we} and van Eijk et al.~\cite{eijk2019impact}, an interesting pattern emerges. As shown in Table~\ref{tab:prev-comp}, although both van Eijk et al.'s and our data collection were conducted after that of Degeling et al., we report lower prevalence than that of the earlier study. This is at odds with the reasonable expectation that the prevalence of cookie banners does not decrease substantially over time. What can explain this discrepancy is the sample size factor. Our results demonstrate that the observed prevalence depends on the size of the sample. That is, although the observed rates might rise initially as samples are expanded from the top hundred to a few hundred websites in each country, further expansion to a few thousands results in a decrease in observed prevalence rates. Hence, our results show that studies with smaller sample sizes might not provide an accurate representation of the big picture. 

\begin{figure}
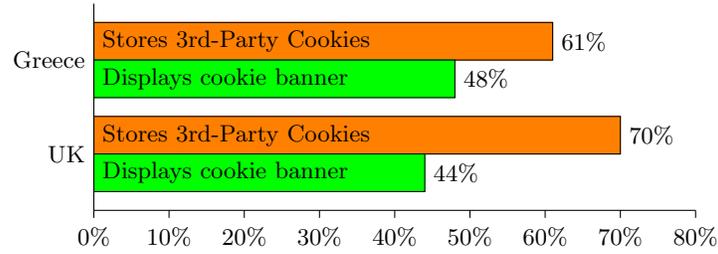

    \centering
    \renewcommand{\bcfontstyle}{}
    \begin{bchart}[step=10,max=80, unit=\%]
        \bcbar[text=Stores 3rd-Party Cookies, color=orange]{61}
        \bclabel{Greece}
        \bcbar[text=Displays cookie banner, color=green]{48}
        \smallskip
        \bcbar[text=Stores 3rd-Party Cookies, color=orange]{70}
        \bclabel{UK}
        \bcbar[text=Displays cookie banner, color=green]{44}
    \end{bchart}
    \caption{Websites that store 3rd-party cookies and display a cookie banner.}
    \label{fig:prevalence_cookie_banners_tps}
\end{figure}

Using the additional data collected by OpenWPM, we found that 61\% of Greek and 70\% of UK websites store at least one third-party cookie on their user’s browser. This suggests that around 13\% of Greek and 26\% UK websites have yet to comply with the GDPR or DPA, respectively, as shown in Fig.~\ref{fig:prevalence_cookie_banners_tps}.

\subsubsection{\ref{rq:avg_options}: Direct opt-outs are rare.}
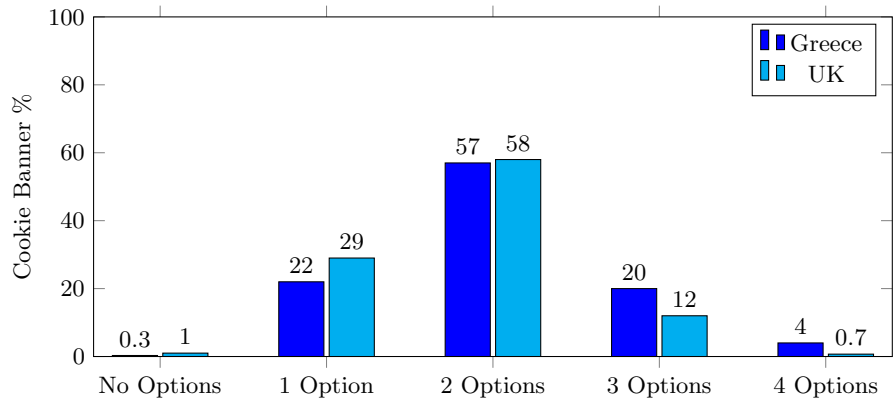
\begin{figure}[t]
    \centering
    \begin{tikzpicture}
        \begin{axis}[
                ybar,
                bar width=.6cm,
                width=\textwidth,
                height=.5\textwidth,
                symbolic x coords={No Options,1 Option,2 Options,3 Options, 4 Options},
                xtick=data,
                nodes near coords,
                nodes near coords align={vertical},
                ymin=0,ymax=100,
                ylabel={Cookie Banner \%},
            ]
            \addplot[fill=blue] table[x=interval,y=Greece]{\categories};
            \addplot[fill=cyan] table[x=interval,y=UK]{\categories};
            \legend{Greece, UK}
        \end{axis}
    \end{tikzpicture}
    \caption{Distributions of cookie banners per number of options in Greece and the UK.}
    \label{fig:avg_options}
\end{figure}

The distribution of the number of options cookie banners in our dataset provide is depicted in Fig.~\ref{fig:avg_options}. 
As the figure shows, the most prevalent number of options in both countries is two. 
The mean number of options is 2.1 for Greece and 1.8 for the UK. 
The median number of options is 2 for both countries. 
This is in agreement with van Eijk et al.'s finding that the median number of choices in the top 100 popular websites that have a cookie banner in both countries was two~\cite{eijk2019impact}. 

\pgfplotstableread[row sep=\\,col sep=&]{
    interval    & Greece & UK \\
    No Options  & 0.3    & 1 \\
    1 Option    & 22     & 29  \\
    2 Options   & 57     & 58 \\
    3 Options   & 20     & 12 \\
    4 Options   & 4      & 0.7 \\
}\categories

\begin{figure}[t]
    \centering
    \begin{tikzpicture}
        \begin{axis}[
                ybar,
                bar width=.6cm,
                width=\textwidth,
                height=.5\textwidth,
                symbolic x coords={Affirmative,Negative,Managerial,Informational},
                xtick=data,
                nodes near coords,
                nodes near coords align={vertical},
                ymin=0,ymax=100,
                ylabel={Cookie Banner \%},
            ]
            \addplot[fill=blue] table[x=interval,y=Greece]{\categories};
            \addplot[fill=cyan] table[x=interval,y=UK]{\categories};
            \legend{Greece, UK}
        \end{axis}
    \end{tikzpicture}
    \caption{The proportion of cookie banners in Greece and the UK providing each type of option.}
    \label{fig:priv_categories_breakdown}
\end{figure}
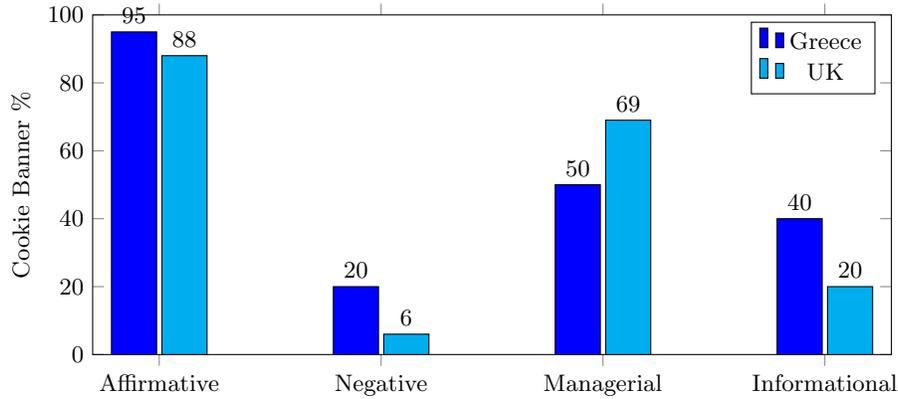

Worryingly, we can see in Fig.~\ref{fig:avg_options} that a considerable proportion of cookie banners provide either no option or only one option to the user. 
This prompts us to look into the distribution of the four categories of privacy options in the cookie banners. 
The results are depicted in Fig.~\ref{fig:priv_categories_breakdown}. 
As the figure shows, although Affirmative options are quite ubiquitous in cookie banners in both countries (Greece: 95\%, UK: 88\%), Negative options are quite rare (Greece: 20\%, UK: 6\%). 
In the upcoming sections we will look further into the exact combinations of options provided by the websites to be able to draw further conclusions. 

\pgfplotstableread[row sep=\\,col sep=&]{
    interval        & Greece & UK    \\
    Affirmative     & 95   & 88  \\
    Negative        & 20   & 6  \\
    Managerial      & 50  & 69 \\
    Informational   & 40  & 20 \\
}\categories

\subsubsection{\ref{rq:no_options}: Most cookie banners nudge towards privacy-intrusive choices.}
Considering all the ways the four categories of options that we have discussed may appear in a cookie banner results in 16 possible combinations. 
We depict the distributions of all these 16 option combinations in both countries in Table~\ref{tab:combos-all}. The combinations are coded with abbreviations in the table, e.g. \texttt{A-M-} stands for the combinations in which at least an Affirmative choice is present, Negative absent, Managerial present, and Informational absent. 

As Table~\ref{tab:combos-all} shows, by far the most prevalent combination is that of Affirmative and Managerial options (i.e. \texttt{A-M-}), with other combinations including an Affirmative option but excluding a Negative option (i.e. \texttt{A-MI}, \texttt{A--I}, and \texttt{A---}) following in terms of prevalence in both countries. 
This shows that at least 75\% of cookie banners in Greece and 82\% in the UK explicitly nudge their users towards accepting cookies. 

Going beyond nudging, as Nouwens et al. argue~\cite{nouwens2020dark}, implicit consent and reject not being as easy as accept are both violations of GDPR and DPA. Let us now consider the 16 combinations against these two criteria. 

For explicit consent, one requires at least an Affirmative or a Managerial option to be present so that the user can register their consent explicitly through one of these options. Hence, all combinations without any of these two options (i.e. \texttt{-N-I}, \texttt{-N--}, \texttt{---I}, and \texttt{----}) represent cookie banners that are violating this criterion. Hence, our results show that at least 16 Greek and 129 UK websites are non-compliant with GDPR and DPA since they do not provide the means for their users to register their explicit consent to the use of cookies. These constitute around 1\% of Greek and 2\% of UK websites with cookie banners. 

The proportions of websites not providing an explicit consent option discussed above are large under-estimations since consent is not necessarily explicit in other combinations. More specifically, in our Affirmative category, apart from terms such as \say{accept} and \say{I agree} that clearly indicate consent, there are many other terms with less clear meaning such as \say{close}, \say{continue}, and \say{dismiss}. These less clear terms roughly constitute around one sixth of all of the observed Affirmative options. We do not believe that such terms are sufficient to indicate explicit consent and hence estimate the level of non-compliance in terms of explicit consent to be around 15\%. 

The situation is much worse if the relative ease of Affirmative and Negative options are considered. Any combination with an Affirmative choice but without a Negative one (i.e. \texttt{A-MI}, \texttt{A-M-}, \texttt{A--I}, and \texttt{A---}) clearly does not provide a negative option as easily accessible as an Affirmative one. Furthermore, any cookie banner that only includes an Informative option or no option (i.e. \texttt{---I} and \texttt{----}) is defaulting on acceptance of cookies if the user navigates away from the cookie banner to interact with the website, hence not providing any means for the user to register their lack of consent. Therefore, all of these combinations do not satisfy the criterion either. This means that overall, our results demonstrate that at least 76\% of Greek and 84\% of UK cookie banners violate the GDPR and DPA in that they do not provide their users with a Negative option as easily accessible as an Affirmative one. 

\begin{table}[t]
    \centering
    \caption{Distribution of the 16 combinations of Privacy Options in Greece and the UK, highlighting those that directly violate the GDPR and the Data Protection Act 2018. \texttt{A}: Affirmative, \texttt{N}: Negative, \texttt{M}: Managerial, \texttt{I}: Informational.}
    \begin{tabular}{@{}c@{\qquad}r@{\quad}r@{\qquad}c@{\quad}c@{}}
    \toprule
        \textbf{Combination} & 
        \textbf{GR} & 
        \textbf{UK} & 
        \textbf{\begin{tabular}[c]{@{}c@{}}Consent \\ Explicit\end{tabular}} & 
        \textbf{\begin{tabular}[c]{@{}c@{}}Accept as easy \\ as Reject\end{tabular}} \\ 
        \midrule
        \texttt{ANMI}   & 4\%       & 1\%       &       &       \\
        \texttt{ANM-}   & 5\%       & 3\%       &       &       \\
        \texttt{AN-I}   & 10\%      & $<$1\%    &       &       \\
        \texttt{AN--}   & 2\%       & 1\%       &       &       \\
        \texttt{A-MI}   & 5\%       & 8\%       &       & No    \\
        \texttt{A-M-}   & 32\%      & 47\%      &       & No    \\
        \texttt{A--I}   & 21\%      & 8\%       &       & No    \\
        \texttt{A---}   & 17\%      & 19\%      &       & No    \\
        \texttt{-NMI}   & $<$1\%    & 0\%       &       &       \\
        \texttt{-NM-}   & $<$1\%    & $<$1\%    &       &       \\
        \texttt{-N-I}   & 0\%       & 0\%       & No    &       \\
        \texttt{-N--}   & 0\%       & 0\%       & No    &       \\
        \texttt{--MI}   & $<$1\%    & 1\%       &       &       \\
        \texttt{--M-}   & 4\%       & 9\%       &       &       \\
        \texttt{---I}   & 1\%       & 1\%       & No    & No    \\
        \texttt{----}   & $<$1\%    & 1\%       & No    & No    \\ 
    \bottomrule
    \end{tabular}
    \label{tab:combos-all}
\end{table}

Degeling et al.~\cite{degeling2018we} report their observations of a several types of cookie banners of top 500 websites in Greece and the UK. Three of their categories can be roughly comparable to collections of combinations we report. 
Cookie banners with \say{no options} in their work roughly correspond to combinations with neither an Affirmative nor a Negative option (i.e. \texttt{--??} where \texttt{?} is a wildcard). They report around 20\% and 40\% for this category (estimated from~\cite[Figure~5(a)]{degeling2018we}) compared to our 5\% and 12\% respectively for Greece and the UK. 
Cookie banners with \say{confirmation only} in their work roughly correspond to combinations with an Affirmative but not a Negative option (i.e. \texttt{A-??}). They report around 65\% and 35\% for this category (estimated) compared to our 75\% and 82\%. 
Cookie banners with a \say{binary} choice in their work roughly correspond to combinations with both an Affirmative and a Negative option (i.e. \texttt{AN??}). They report around 4\% and 5\% for this category (estimated) compared to our 20\% and 5\%. 
These comparisons show that observed practices may substantially vary between observations of smaller and larger sample sizes. 

Limiting their attention to cookie banners provided by the 5 most popular CMPs in the UK, Nouwens et al. found around 75\% violating the \say{reject as easy as accept} criterion~\cite{nouwens2020dark}. 
Our analysis gives the rate of at least 84\% for the violation of this criterion showing that the situation is much worse when a larger set of websites are considered. 

In addition to privacy options, cookie banners usually contain a concise textual description as well. The text's primary function is to inform users why cookies are used and how they may affect them. This text is usually considerably shorter compared to the full Privacy Policy of the website. 
Examples of such texts can be seen in Fig.~\ref{fig:cookie_banners}.

The average length of cookie banner texts in Greek websites was 66 words, slightly longer than the UK average of 52 words. 

Employing the term frequency--inverse document frequency (TF-IDF) formula to identify the most prominent terms in the cookie banner text corpus, we found that the most prominent terms in Greece and the UK are quite similar and dominated by terms with an apparent positive connotation such as \say{best}/\say{better}/\say{$\kappa \alpha \lambda \acute{\upsilon} \tau \varepsilon \rho \eta$}, \say{ensure}, and \say{experience}/\say{$\varepsilon \mu \pi \varepsilon \iota \rho \acute{\iota} \alpha$}. 

In fact, none of the top 50 prominent terms in either country (available from the repository) appear to have a negative connotation, whereas terms with a positive connotation such as \say{improve}/\say{$\beta \varepsilon \lambda \tau \iota \acute{\omega} \sigma \varepsilon \iota$} and \say{enhance} constitute a considerable proportion of the list of terms. 

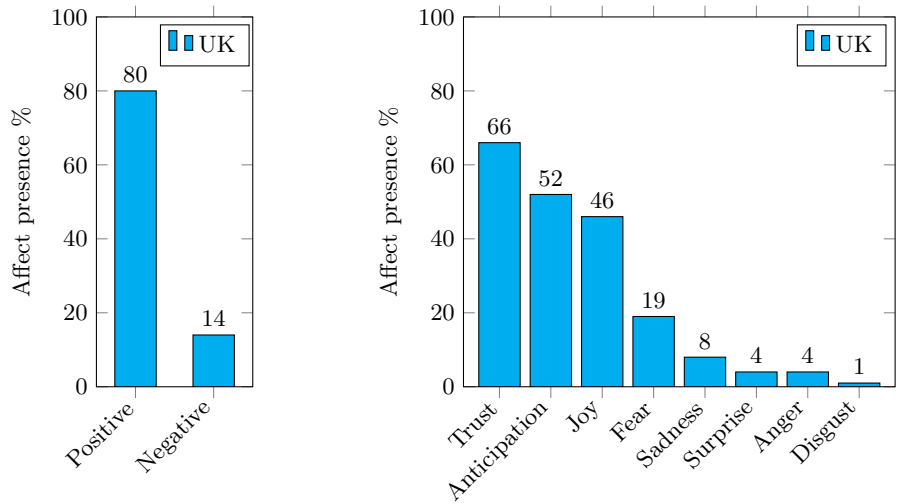
\begin{figure}[t]
    \centering
    \begin{minipage}[t]{0.3\textwidth}
        \begin{tikzpicture}[baseline=(current bounding box.north)]
            \begin{axis}[
                    ybar,
                    bar width=.55cm,
                    width=\textwidth,
                    height=20em,
                    symbolic x coords={Positive, Negative},
                    xtick=data,
                    nodes near coords,
                    nodes near coords align={vertical},
                    ymin=0,ymax=100,
                    ylabel={Affect presence \%},
                    x tick label style={rotate=45,anchor=east},
                    enlarge x limits=0.5
                ]
                \addplot[fill=cyan] table[x=Affect,y=UK]{\ukpol};
                \legend{UK}
            \end{axis}
        \end{tikzpicture} 
    \end{minipage}
    \hfill
    \begin{minipage}[t]{0.6\textwidth}
        \begin{tikzpicture}[baseline=(current bounding box.north)]
            \begin{axis}[
                    ybar,
                    bar width=.55cm,
                    width=\textwidth,
                    height=20em,
                    symbolic x coords={Trust, Anticipation, Joy, Fear, Sadness, Surprise, Anger, Disgust},
                    xtick=data,
                    nodes near coords,
                    nodes near coords align={vertical},
                    ymin=0,ymax=100,
                    ylabel={Affect presence \%},
                    x tick label style={rotate=45,anchor=east},
                    % enlarge x limits=0.2
                ]
                \addplot[fill=cyan] table[x=Affect,y=UK]{\ukaff};
                \legend{UK}
            \end{axis}
        \end{tikzpicture}
    \end{minipage}
    \caption{Distributions of the emotional affects in the UK cookie banner text corpus}
    \label{fig:sentiment-analysis}
\end{figure}

To get a more comprehensive view of the connotations relayed by cookie banner texts in the UK, we performed an automated sentiment analysis of the words used in all UK banner texts using NRCLex~\cite{nrclex}. 
The analysis found that a generally positive emotional affect was present in around 80\% of the banner texts, whereas a generally negative affect was present in only around 14\%. 
Besides, an overwhelming majority (of more than 9 in 10) of the texts with a negative affect also had a positive effect present as well. 
Looking at more specific emotional affects, trust and joy are among the most prevalent, present in around 66\% and 46\% of the texts, respectively. 
The prevalence of general and specific emotional affects is shown in Fig.~\ref{fig:sentiment-analysis}. 

\pgfplotstableread[row sep=\\,col sep=&]{
    Affect          & UK \\
    Positive        & 80 \\
    Negative        & 14 \\
}\ukpol

\pgfplotstableread[row sep=\\,col sep=&]{
    Affect          & UK \\
    Fear            & 19 \\
    Anger           & 4 \\
    Anticipation    & 52 \\
    Trust           & 66 \\
    Surprise        & 4 \\
    Sadness         & 8 \\ 
    Disgust         & 1 \\
    Joy             & 46 \\
}\ukaff

The automated term prominence and sentiment analyses above suggest that websites tend to give a one-sided description of cookie usage, namely that it enhances browsing experience, conveniently leaving out that cookies can be used for tracking. This is in line with previous manual analyses of smaller samples, e.g.\ that of Utz et al.~\cite{utz2019informed}, that found similar biases in cookie banner texts.

\subsubsection{\ref{rq:manage_options_count}: Managing trackers is more prevalent than opting out.}
We aimed to determine whether websites allow their visitors to manage their privacy settings from the cookie notice. Our results show that Managerial options in cookie banners are significantly more prevalent compared to the Negative ones (See Fig.~\ref{fig:priv_categories_breakdown}). More specifically, 59\% of Greek and 69\% of UK cookie banners offer a Managerial option compared to 20\% and 6\%, respectively, with a Negative one. Hence, users in both countries are several times more likely to be given an option to manage their cookies than an option to decline them. 

\subsubsection{\ref{rq:distribution}: Users in both countries lack real choice, but practices vary.}
The results discussed in the previous sections show that Greek and UK users face a largely similar landscape in terms of the prevalence of cookie banners, widespread deployment of third-party cookies, and rampant use of nudging and lack of consent to cookies being much harder to register than consent. 
However, there are some notable differences between the two countries. 
With respect to using third-party cookies but not showing a cookie banner at all, the proportion of websites that are non-compliant with regulations in the UK is almost double that in Greece. Besides, cookie banner in Greece tend to provide slightly higher number of options. 

Looking at the option types, affirmative options are prevalent in both, but negative options are much scarcer in the UK. From the other two types, managerial options are more prevalent in the UK and informational ones more in Greece. Looking at the option combinations, most have similar prevalence with three exceptions: although \texttt{AN-I} can be found in around 10\% of banners in Greece, it is very rare in UK; similarly, \texttt{A--I} has a much higher share in Greece (21\%) than in the UK (8\%); on the other hand, \texttt{A-M-} is found in almost half of the cookie banners in the UK, but only in about a third in Greece. 

\section{Conclusion}
We set out to conduct the most comprehensive study of cookie banners in the UK and Greece to date in the hope that a more thorough understanding of the cookie banner landscape in the two countries is beneficial for a rage of stakeholders including users, privacy-enhancing technology developers, and policymakers.

By extending OpenWPM to detect and store cookie banners, over 17,000 websites were crawled and more than 7,000 cookie banners were collected. 

Our results show that although around half of the websites in our dataset display a cookie notice, a substantial proportion do not show one even though they use third-party cookies. 
Furthermore, websites make it extremely difficult for users to opt-out from tracking with only a minority offering a direct opt-out option. 
Our analysis also suggests that websites present cookies as devices that improve browsing experience for the user while the negative aspects of tracking tend to be downplayed. 

Hence, we find clear evidence of websites nudging visitors towards privacy-intrusive choices and violating regulations. 

Although in many cases our results agree with previous studies considering smaller samples, we also found that in some cases, e.g. prevalence of cookie banners and those providing specific options, our observations significantly differ from previous reported values. 
Hence, we hope that our work provides a more holistic view of the landscape of cookie banners in the two countries. 

Future work directions include more comprehensive studies of the cookie banner landscape for other countries (for which our code is available and can be reused), a more detailed analysis and classification of varying cookie banner practices in specific subsets of the dataset, e.g., in different industries, and further analyses of the cookie banner text corpus. 

\bibliographystyle{splncs04}
\bibliography{Bibliography}

\end{document}